\documentclass{article}
\usepackage[utf8]{inputenc}
\usepackage{authblk}
\usepackage{hyperref}
\usepackage{booktabs}
\usepackage{graphicx}

\title{Analyzing Hack Subnetworks in the Bitcoin Transaction Graph}
\author[1]{Daniel Goldsmith\thanks{daniel.goldsmith@chainalysis.com}}

\author[1]{Kim Grauer\thanks{grauer@chainalysis.com}}

\author[1]{Yonah Shmalo\thanks{yonah.shmalo@chainalysis.com}}

\affil[1]{Chainalysis}
\date{October 2019}

\begin{document}

\maketitle
\begin{abstract} 

Hacks are one of the most damaging types of cryptocurrency related crime, accounting for billions of dollars in stolen funds since 2009. Professional investigators at Chainalysis have traced these stolen funds from the initial breach on an exchange to off-ramps, i.e. services where criminals are able to convert the stolen funds into fiat or other cryptocurrencies. We analyzed six hack subnetworks of bitcoin transactions known to belong to two prominent hacking groups. We analyze each hack according to eight network features, both static and temporal, and successfully classify each hack to its respective hacking group through our newly proposed method. We find that the static features, such as node balance, in degree, and out degree are not as useful in classifying the hacks into hacking groups as temporal features related to how quickly the criminals cash out. We validate our operating hypothesis that the key distinction between the two hacking groups is the acceleration with which the funds exit through terminal nodes in the subnetworks. \end{abstract}
\providecommand{\keywords}[1]
{
  \small	
  \textbf{\textit{Keywords---}} #1
}

\section*{Introduction}

The Bitcoin network  is a distributed, public ledger, secured through blockchain technology. All transactions occur between two distinct public addresses and are permanently recorded on the specific blockchain built for bitcoin. The process of securing these transactions is handled by bitcoin miners, who use their computing power to solve complex cryptographic problems and in the process verify blocks and transactions \cite{Nakomoto}. 

Anyone can create a bitcoin address to receive funds through a variety of software projects such as Blockchain.info \cite{blockchain.info} or Electrum wallets \cite{electrum}. Additionally, there is no limit to the number of bitcoin addresses that any individual or organization can make. There are also no requirements for verifying your identity in the process of address creation. It is completely free to make an address, however, it costs money to transfer money on the network by paying transaction fees. 

Because of the ease of transactions between pseudonymous addresses, cryptocurrencies, and bitcoin in particular have been especially attractive to criminals who both exploit technological vulnerabilities and prefer to move funds through the pseudonymous bitcoin transaction network to avoid detection by law enforcement \cite{Ransomware}. Indeed, the amount of cybercrime involving cryptocurrencies has grown via ransomware \cite{Ransomware}, scamming activity, phishing scams, and hacking of exchanges or wallets \cite{cryptocrime}.

Notably, exchange hacks are one of the most costly types of cryptocurrency related crime. Hackers have stolen $\$$1.7 billion dollars worth of cryptocurrency from exchanges since 2011 \cite{cryptocrime}. Tracing stolen funds in order to freeze the assets of the perpetrators is one of the most effective ways of safeguarding against future attacks, as this method removes bad actors from the ecosystem and disincentivizes similar activity from other actors. Typically, either government or private cyberinvestigators, take up the task of tracing stolen cryptocurrency funds. Their investigations begin with a known address that has been hacked. They then follow the funds through up to thousands of different addresses until the funds hit a service (an off-ramp), i.e. an alternative means of cashing out the stolen bitcoin. Ideally, an investigator will trace funds to a service so that a subpoena can be issued to the service to unmask the identity of the criminal. These investigations result in traced out subnetworks representing the flow of stolen bitcoin from the point of breech on an exchange through exit ramps. The subnetworks analyzed here were provided by investigators at Chainalysis, a firm specializing in blockchain investigations.

We present research to algorithmically visualize and analyze hack cash out subnetworks that capture the temporal behavior of hackers and locate the stolen funds. We then build similarity matrices based on eight graph features, run community detection over those matrices, and successfully classify certain hacks to the known hacking organization to have carried out the attack. We find that temporal features, such as the rate at which the hackers send funds to exit ramps, are the most effective features to use for grouping specific hacks together and classifying them to their hacking groups.

\subsection*{Algorithmically traversing hack subnetworks and its limitations}

We investigate bitcoin hacks by traversing subnetworks of nodes that have been built out by professional crime investigators. These hack subnetworks are comprised of nodes that have either directly or indirectly received hacked funds.

\begin{figure}[htp]
\centering
    \includegraphics[width=8cm,height=6cm]{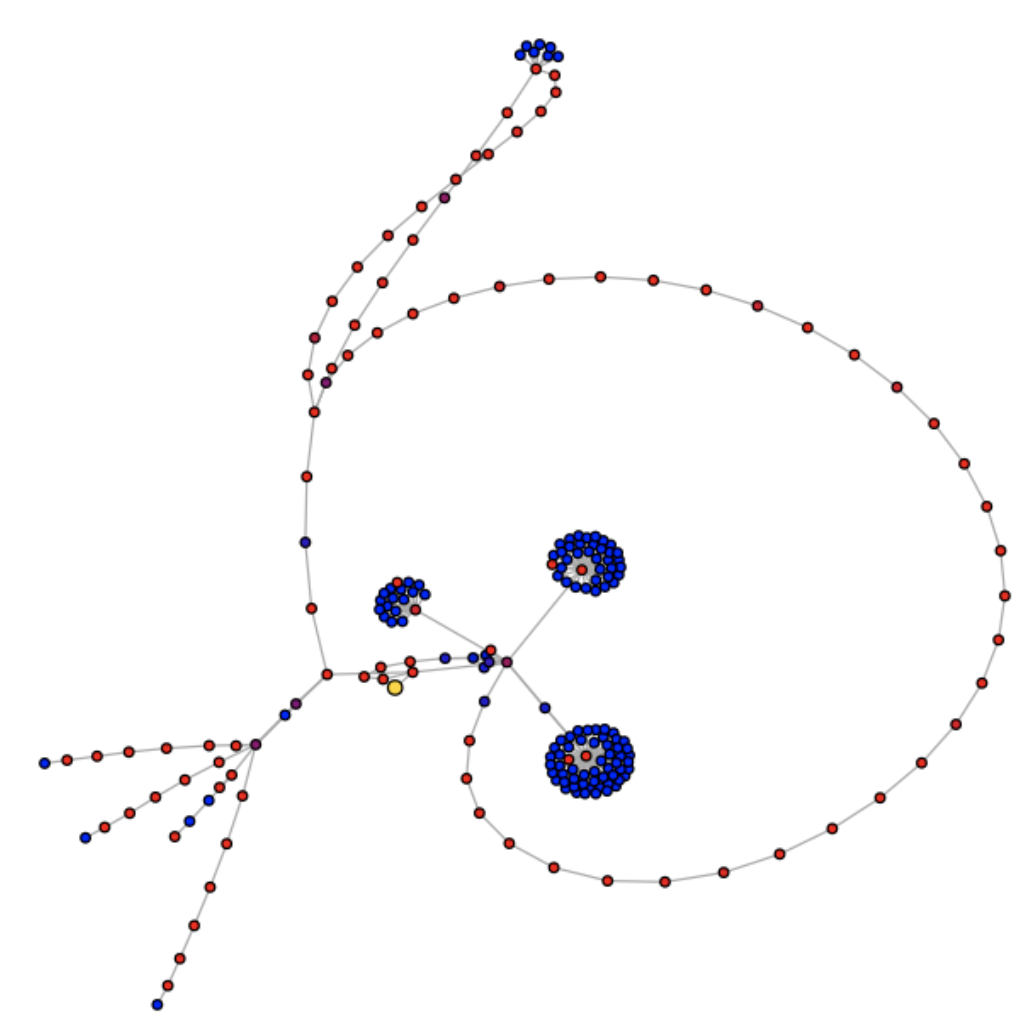}
    \caption{Sample Hack Subnetwork, Hack A3}
    \label{networkgraph}
\end{figure}

We then create visualizations to identify trends in the hack and to better understand the time patterns specific to each hack as the stolen bitcoin flows to terminal nodes, see Figure \ref{amountinplayovertime}. In some cases, when the level of obfuscation is minimal, investigations tracking stolen funds often terminate at services (see Methodology section on identifying services), simply because criminals want to change their stolen bitcoin for fiat currency, or at least convert it to a another cryptocurrency. 

Yet cryptocurrency investigations are usually much more complex then this \cite{nouh}. Often, the investigator may not know if a node belongs to a service, particularly in the case of a mixing service. Furthermore, stolen bitcoin from some of the largest hacks may utilize laundering mechanisms in which OTC brokers act as third party sellers allowing for a change of hands to an entity that is no longer behind the hack. This activity can not be detected through blockchain analytics unless their is a source of ground truth confirming the funds passing through on OTC broker. Without this confirmation, the funds would appear to move from one pseudonymous node to another. 

Sometimes the investigations are so complex that the investigator simply cannot go through the process of tracing every single stolen bitcoin to an cash out point. In this case, the investigator may choose to chase particularly promising leads, rather then spend the time to analyze every single transaction that occurred. Additionally, the actual concept of a terminal node may be less clear. At any given time, stolen funds may be sitting idly in non-service clusters for extended periods of time. In practice, it is common for funds to slowly leak out of these ``holding" clusters \cite{cryptocrime}. In this case, a node is neither clearly a pass through address nor a terminal node. 

In all of these cases, using the actual hack subnetwork may be insufficient for conducting a comprehensive analysis to determine the underlying trends. We therefore develop the concept of \textit{terminal nodes} to monitor funds leaking out through both known exit points, such as identified services, and unknown exit ramps, which are candidates for terminal nodes. 

Because the graphs have such limitations, we develop a parameter, $\rho$, to define terminal nodes based on the ratio of sending to receiving activity for a cluster. The $\rho$ parameter can be set by the investigator as a way to modify the natural edges of the graph. Continuously varying the $\rho$ parameter from 0\% to 100\%, and observing the subsequent changes in the stolen funds flowing to terminal nodes, is also an effective method upon which we build our pipeline in the following section.

\section*{Methodology}

\subsection*{Pipeline}

\begin{figure}[htp]
\centering{}
    \includegraphics[width=8cm,height=6cm]{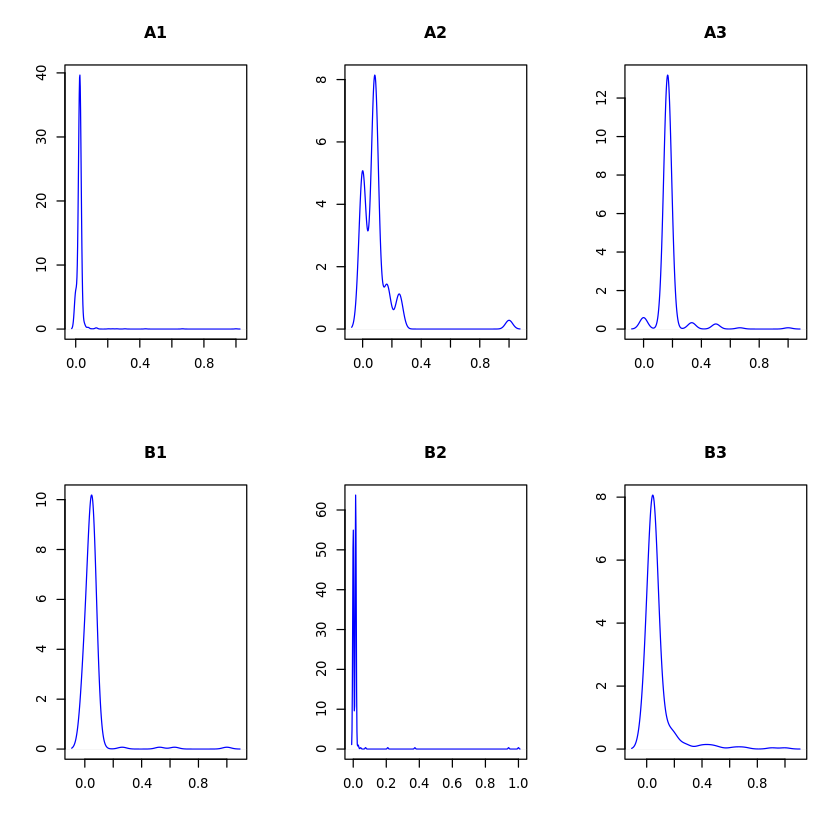}
    \caption{Distribution of In Degree}
    \label{indegree}
\end{figure}

\begin{figure}[htp]
\centering{}
    \includegraphics[width=8cm,height=6cm]{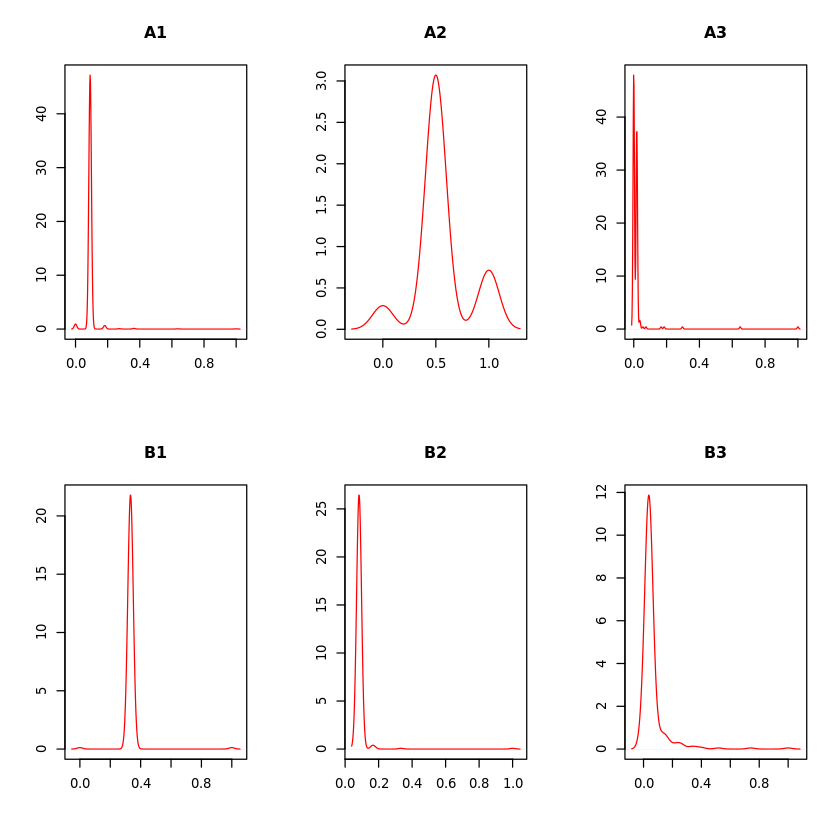}
    \caption{Distribution of Out Degree}
    \label{outdegree}
\end{figure}

\begin{figure}[htp]
\centering{}
    \includegraphics[width=8cm,height=6cm]{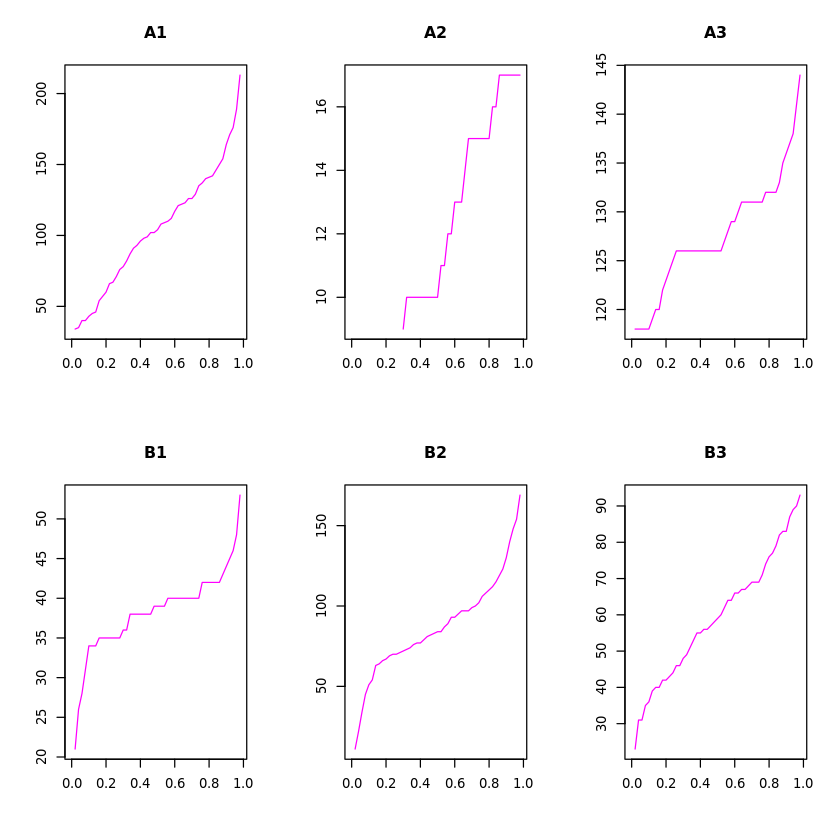}
    \caption{Number of Terminal Nodes as function of $\rho$, TvR}
    \label{terminalnodesvsrho}
\end{figure}

\begin{enumerate}
  \item We first gather subnetworks of known hacks that have been built out by professional investigators.
  \begin{itemize}
 \item Due to the sensitivity of this data and relative infrequency of hack events, the result of this process provided a small set of anonymized, curated subnetworks that trace stolen funds from the origin of the hacks to all end points of interest.
 
  \item It is at this point that we introduce a new tool for analyzing these subnetworks for additional insights that we can eventually return to the investigators and compliance officers at exchanges.
  \end{itemize}
  \item We traverse these subnetworks from the starting clusters until the funds have been fully cashed out at exit points, i.e. terminal nodes.
   \begin{itemize}
            \item An element of complexity emerges in this analysis that requires additional attention, namely that the terminal nodes require a more rigorous definition than any cluster sitting on the outskirts of the subnetwork since many of these terminal nodes act as sinks but still slowly leak funds despite maintaining control over the majority of their hacked balance.
    \end{itemize}
  \item Next, to better visualize the temporal activity in the hacks, we create two time series that display the activity of the hacked funds.
    \begin{itemize}
        \item First, we measure how active the hackers are over time by computing the number of transfers the hackers make each day, as seen in Figure \ref{transfersovertime}.
        \item Second, we measure the funds traced as they move to terminal nodes, as seen in Figure \ref{amountinplayovertime}. As the funds move through terminal nodes, the share of funds still held by hackers decreases. A fully tracked hack subnetwork would be visualized by the funds decreasing from 100\%\ to 0\%\ of funds still held by the hacker over the number of days that it takes to fully exit the funds through terminal nodes.
    \end{itemize}
    \item We then generate distributions for the following features for each hack subnetwork:
    \begin{itemize}
        \item Logarithm of Hack balance of all nodes, see Figure \ref{Distributionlogbalance}.
        \item Weighted In-degree of all nodes, see Figure \ref{indegree}.
        \item Weighted out-degree of all nodes, see Figure \ref{outdegree}.
        \item Logarithmic difference of the average percent of funds still in play, across all $\rho$ values, derived from data shown in Figure \ref{amountinplayovertime}.
        \item Second difference of the average percent of funds still in play, across all $\rho$ values, derived from data shown in Figure \ref{amountinplayovertime}.
        \item Logarithmic difference of the standard deviation of the percent of funds still in play, across all $\rho$ values, derived from data shown in Figure \ref{stdovertime}.
        \item Average number of transactions to terminal nodes per day, across all $\rho$ values, derived from data shown in Figure \ref{transfersovertime}.
        \item Terminal Nodes as a function of $\rho$, see Figure \ref{terminalnodesvsrho}.
    \end{itemize}
    \item Afterwards, we create similarity matrices corresponding to each distribution, whose elements are the pairwise similarities of the distributions corresponding to each of the hack subnetworks via the 1-Dimensional Wasserstein Distance, i.e. the Earthmover Distance \cite{wasserstein1d}\cite{encyclopediaofmath}.
    \item We run two community detection algorithms, Modularity Optimization \cite{Modularity}\cite{modpaper} and Walktrap \cite{Walktrap}\cite{walkpaper}. We compare the output of the overall approach across the similarity matrices for all the distributions against our ground truth attribution of the two underlying hacking groups and demonstrate the potential for such a method by properly reattributing the hack networks to their respective groups. 
    \item Lastly, we review the output communities and test our hypothesis that the features relating to the hack dynamics are more informative in classifying the hacking groups than the static network features.
\end{enumerate}

\subsection*{Identifying Services}

A typical service can control thousands of addresses, while larger services can even manage into the millions. We identify services by exploiting features unique to the Bitcoin blockchain. There are many different approaches that blockchains employ to cryptographically verify transactions, but the Bitcoin blockchain relies on Unspent Transaction Outputs (\textit{UTXO's}) to record all transactions. A UTXO is the unspent output of a previous transaction that a user is entitled to transfer to another bitcoin address. Every wallet that holds a positive bitcoin balance is in possession of at least one UTXO. When multiple UTXO's are held by a single user and spent together in a transaction, it then becomes possible to definitively ascribe common ownership to all of the UTXO's that were spent together. This concept of a \textit{cospend} is the basis of the clustering activity used by blockchain analysis firms such as Chainalysis to identify clusters of addresses controlled by a single entity. The network then becomes comprised of cospend clusters, i.e. nodes, composed of multiple addresses rather than long chains of single-use addresses \cite{meiklejohn}. 

Once addresses have been mapped to a node through cospending activity, the node can be mapped to a named entity by interacting directly with it. For the example of an exchange, this process can occur by visiting an exchange's website, depositing funds on the exchange, and tracing that transaction via a block explorer \cite{blockchain.info}. Only services with publicly available address information can be identified in this way.

When stolen funds arrive at a known service, such as a an exchange, we can assume that the hackers have attempted to cash out their funds. Professional investigators trace funds through these nodes to create hack subnetworks that capture as much of the meaningful movement of the stolen funds as possible.

\subsection*{Defining Terminal Nodes}

There are four types of terminal nodes discussed in this paper. 1) A known service terminal node that is a confirmed service through the process mentioned above of pairing ground truth knowledge with cospending activity. These services can be exchanges, mixers, gambling sites, merchant service platforms, or any exit ramp through which a criminal can off-load stolen bitcoin to an institutional cryptocurrency player. 2) An unknown service node, where the investigator has reason to believe a node is behaving like a service and will therefore terminate the investigation at that point, 3) a node that the investigator has no reliable information for, or 4) a node in which the investigation terminates because the investigator decided not to pursue the lead. 

By default, terminal nodes are the edges of the graph subnetwork. Ideally, a subnetwork of a hack would track 100\% of the funds from the point of a hack through all exit ramps. This would allow us to set $\rho=0.00$, as the terminal nodes would simply be all the natural edges of the graph. In this case, the investigator would trace funds to a service, whether it be an exchange, mixing site, gambling site, etc. $\rho=0.00$ indicates that a node has only ever received funds within the subnetwork.

We focus on the ratio rather than the difference of funds sent to received because we want to maximize the number of meaningful leads for investigators rather than raw amount due to hacked funds. By returning this normalized list of terminal nodes and resulting charts, we find all partial sinks ``of interest" in the subnetwork that may facilitate the issuance of subpoenas or other leads, as well as wallets to watch because they still contain funds, large or small. As a secondary filter, we can sort by balance due to the hack, but this feature is only relevant in the operational stage for investigators, not when conducting our analysis.

We define $\rho$ as:

\vspace{\baselineskip}

{\centering
  $\rho = \frac{\textit{weighted in-degree}}{\textit{weighted out-degree}}$, 
\par
} 
\vspace{\baselineskip}

i.e. the ratio of the amount of funds sent to received.

Others have proposed using ratios of the in/out degrees when studying the Bitcoin Transaction Graph, but in different contexts and not as a node-level feature \cite{tessone}.

\begin{figure}[htp]
    \centering
    \includegraphics[width=8cm]{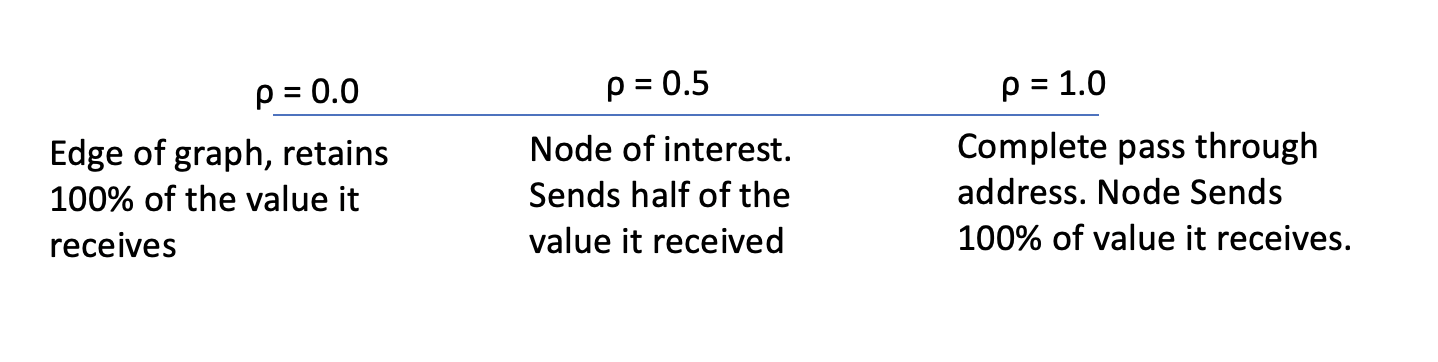}
    \caption{Spectrum of rho values and their significance}
    \label{rho interpretation}
\end{figure}

 Figure \ref{rho interpretation} shows the spectrum of $\rho$ values and their subsequent interpretation. \newline

\subsection*{Visualizing temporal behavior in the hack subnetworks}

The temporal visualizations are shown in Figures \ref{transfersovertime} and \ref{amountinplayovertime}. Figure \ref{transfersovertime} shows the number of transfers over time within the hack subnetwork so that the investigator can get a sense of how active the hackers are over time. They can answer questions such as: does the hacking group consistently make transactions over time, or do they tend to move funds according to a temporal pattern. A pattern may be indicative of an algorithm moving the funds, as opposed to actual individuals approving the transactions.

Figure \ref{amountinplayovertime} shows how the funds exit over time through terminal nodes. It allows an investigator to see the exiting strategy of the hacking group in time. For example, do the hackers exit the funds in one period of time, or consistently over a longer duration of time? Each of these strategies has implications for how the investigator profiles the hacking group overall. For example, a hacking group that exits all the funds through one exchange in one day may be less organized and less well-funded than a hacking group that gradually, through thousands of strategic transactions, exits the funds over a long period of time.

The trends are made visible by restructuring the hack subnetworks into time series. Figure \ref{transfersovertime} demonstrates how active the hackers are by using the number of transactions they carry out as proxies. 

Figure \ref{transfersovertime} allows us to see the way the hackers utilize terminal nodes. Hacking group alpha (A1) is much more active, slowly moving funds through terminal nodes over a shorter period. Hacking group beta (B1) utilizes fewer transactions in general, but tends to send all of their transfers to terminal nodes in a short period of time. In the case of chart B1 in Figure \ref{transfersovertime}, the hackers sat on their funds for a long period of time before abruptly exiting over 70\% of the funds through a few exit ramps within a one week period.

To test the hypothesis that the hackers are best classified using temporal features such as the rate at which funds cash out at terminal nodes, we vary $\rho$ in the following sensitivity analysis section to observe stolen bitcoin exiting through terminal nodes under a range of conditions.

\subsection*{Sensitivity Analysis of $\rho$}

We allowed $\rho$ to range from 0.02 to 0.98 to test the implications of gradually change the $\rho$ parameter. A cluster with a very low $\rho$ value, e.g. $\rho = 0.1$, would have to hold on to more 90\%\ of the funds it received to be considered a terminal node. On the other hand, a very high $\rho$ value, e.g. $\rho = 0.9$, allows a cluster to retain only 10\%\ of the funds it received from the hack in order for it to be considered a terminal node. A higher $\rho$ will capture many more terminal nodes, as it is an easier condition for nodes to meet.

A lower $\rho$ value means that the there are fewer terminal nodes picked up in the graph, and the criteria for being ``of interest" to an investigator is extremely high.  A very low $\rho$ specifies that wallets of interested are those which may only hold small amounts of the total funds that it received. A node holding over 90\%\ of the funds might be a holding wallet gradually leaking out funds, it might be a consolidation wallet for a criminal ring, a wallet associated with other types of criminal activity, or even a point of conversion to another cryptocurrency if, for example, the wallet is an Exodus wallet, which allows for wallet level cryptocurrency conversions. 

Choosing the right value for $\rho$ allows us to optimally grow the hack subnetwork such that it would include the paths of interest without becoming too large to meaningfully analyze. We found that setting the ratio too high resulted in a less meaningful yet larger hack subnetwork, where the terminal nodes did not adequately capture dynamics of interest, and setting the ratio to be too low did not include clusters that likely should have been included. 

Applying a range of $\rho$ from $\rho=0.02$ through $\rho=0.98$, in increments of $0.02$, had very large implications for the amount of funds considered to be tracked. While changing  $\rho$ typically revealed how much of the funds the investigator tracked, at the same time, changing the $\rho$ value does not impact the overall cash out trend witnessed by the investigator. 

These results indicate that varying $\rho$ may not be useful for understanding the behaviors of the hacker, but is a useful tool for identifying nodes of interest that could be possible leads to the investigator. Indeed the variance in the $\rho$ parameter proved one of the most useful tools for running community detection.

We finally then needed to handle the introduction of funds at a time later than the hack by either the same or different user. To account for this, we either add these new flows to the funds at the start and work with the new total as our amount of hacked funds, or we incorporate these flows into our $\rho$ definition, by stating a further constraint that if $\rho > 1$, then it is a terminal node and we do not follow its flows forward in time. In the case of the former, we can track all funds engaged in clearly illicit activity, regardless of source, while in the case of the latter, we are actively restricting the subnetwork to funds that explicitly originated from the source of the hack.

\subsection*{Feature Definitions}
The goal when selecting which distributions to analyze was to capture the behavior of movement of the hacked funds in a precise way. To confirm the hypothesis that the two hacking groups exhibit different cashout strategies, we decided to consider the empirical distributions of 8 different features, as mentioned in Step 4 of the Pipeline. We define several of the features in our analysis as follows:
\begin{enumerate}
    \item Hack balance of all nodes. 
    \vspace{\baselineskip}
    \begin{center}
        $\textit{Bal} = \log(\textit{weighted in-degree $-$ weighted out-degree}$)
    \end{center} 
   \vspace{\baselineskip}
    \item Logarithmic first difference of the average, \textit{LDA}, percent of amounts still in play, \textit{AIP}, across all $\rho$ values.
    \begin{center}
    \vspace{\baselineskip}
        $\textit{LDA} = \log ( \frac{\mathbf{E}[\textit{AIP}(t+1)]}{\mathbf{E}[\textit{AIP}(t)]})$
    \end{center}
\vspace{\baselineskip}
    \item Second difference of AIP, across all $\rho$ values.
    \begin{center}
    \vspace{\baselineskip}
        $\textit{Second Diff(AIP)} = \frac{LDA(t+1) - LDA(t)}{LDA(t)}$
    \end{center}
  \vspace{\baselineskip}
    \item Logarithmic difference of the standard deviation, \textit{LDST}, of the \textit{AIP}, across all $\rho$ values. \vspace{\baselineskip}
    \begin{center}
        $\textit{LDST} = \log ( \frac{\mathbf{E}[(\textit{AIP}(t+1)-\mathbf{E}[\textit{AIP}(t+1)])^2]}{\mathbf{E}[(\textit{AIP}(t)-\mathbf{E}[\textit{AIP}(t)])^2]})$
    \end{center}
     \vspace{\baselineskip}
    \item Average number of transactions to terminal nodes, \textit{TTN} per day, across all $\rho$ values.
   \vspace{\baselineskip}
    \begin{center}
    $\textit{Transactions} =  \mathbf{E}[TTN]   $
    \end{center}
   \vspace{\baselineskip}
\end{enumerate}
        
\subsection*{Similarity Matrices}
Once all of the normalized histograms were generated, we measure the pair-wise similarity between them, per variable, via the 1-Dimensional Wasserstein Distance, a.k.a. the Earthmover Distance or $L^1$ Norm. Generally, the $L^p$ Norm is defined as: \vspace{\baselineskip}
\begin{center}
    $W_p(F,G) = (\int_0^1 |F^{-1}(u)-G^{-1}(u)|^p \; du)^{1/p},$
\end{center}
\hfill\break
where $F$ and $G$ are empirical distribution functions with generalized inverses, $F^{-1}$ and $G^{-1}$
\cite{encyclopediaofmath}.
\subsection*{Community Detection}
After the similarity matrices are computed for the distributions of interest, the goal becomes differentiating between the two hacking groups. We propose a method of representing the similarity matrices as networks and searching for two distinct communities via both Modularity Optimization and Walktrap and comparing the results. 

Modularity Optimization \cite{modpaper}  consists of finding a near maximal value for Modularity, $Q$, returned from the communities applied to some null model of network formation, typically a Random Network.
\vspace{\baselineskip}
\begin{center}
    $Q = \frac{1}{2m}\sum\limits_{vw}[[A_{vw} - \frac{k_vk_w}{2m}]\delta(c_v,c_w)],$
\end{center}
where $m$ is the number of edges in the network, $A_{vw}$ is 1 when nodes $v$ and $w$ are connected and 0 otherwise, $k_v$ is the sum of $A_{vw}$ over $w$, and $\delta(i,j)$ is 1 when $i$ and $j$ are equal and 0 otherwise.
\newline
Walktrap \cite{walkpaper} operates similarly, also attempting to optimize the same modularity, but with a focus on short random walks exiting communities as the explicit motivation and approach.

Both algorithms are built for analyzing large networks, and their true modularity optimization functions are not explicitly the $Q$ written above, but a derived form. 

We utilized both methods as independent confirmation rather than any benefits from their relative optimizations. As the resulting networks are small, with one node corresponding to each hack, are eight distributions analyzed, and two applications of community detection, any conclusions drawn from our method are only tentative since no conclusive results can be drawn from such small amounts of data. Nevertheless, we propose the full method as technically sound and a novel tool in the analysis of hack subnetworks in the bitcoin blockchain.

\section*{Results}

\begin{table}[ht]
  \caption{Summary Statistics for Each Hack }
  \label{tab:freq}
  \begin{tabular}{cccccc}
    \toprule
    Hack & Txs & Nodes & Avg In Deg& Avg Out Deg & Clust. Coeff\\
    \midrule
    A1 & 1,981 & 1,257 & 1.02 & 1.02 & 0.001\\
    A2 & 421 & 55 & 1.11 & 1.11 & 0.041\\
    A3 & 607 & 218 & 1.05 & 1.05 & 0.008\\
    B1 & 190 & 176 & 1.01 & 1.01 & 0.000\\
    B2 & 374 & 335 & 1.06 & 1.06 & 0.002\\
    B3 & 57,299 & 174 & 1.62 & 1.62 & 0.068\\
  \bottomrule
\end{tabular}
\end{table}

\begin{figure}[p]
\centering{}
    \includegraphics[width=8cm,height=6cm]{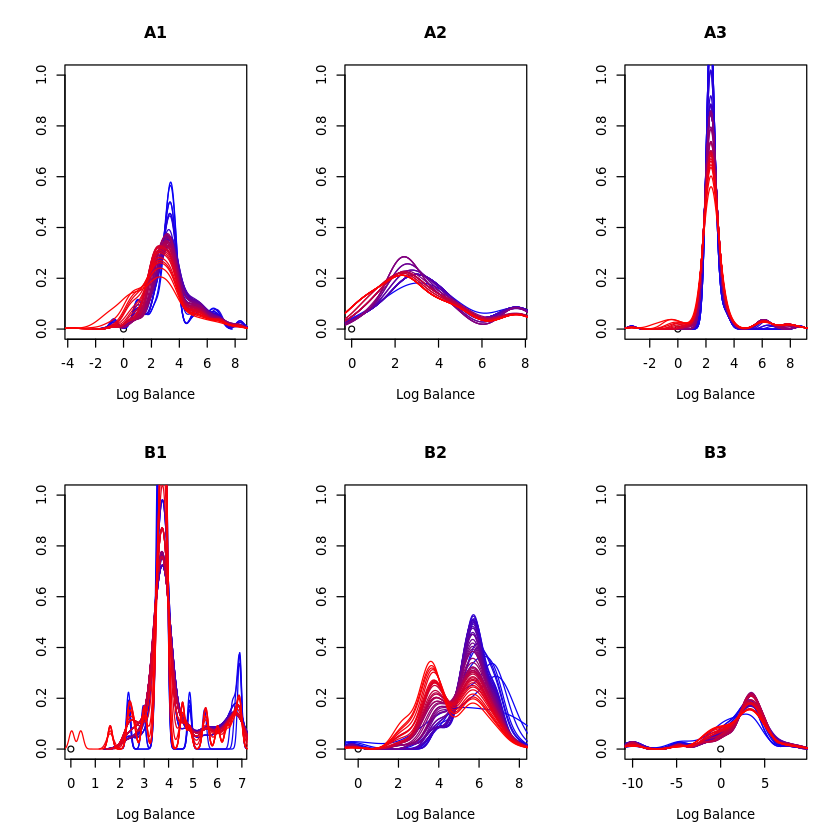}
    \caption{Distribution of Log Balance}
    \label{Distributionlogbalance}
\end{figure}

\begin{figure}[htp]
\centering{}
    \includegraphics[width=8cm,height=6cm]{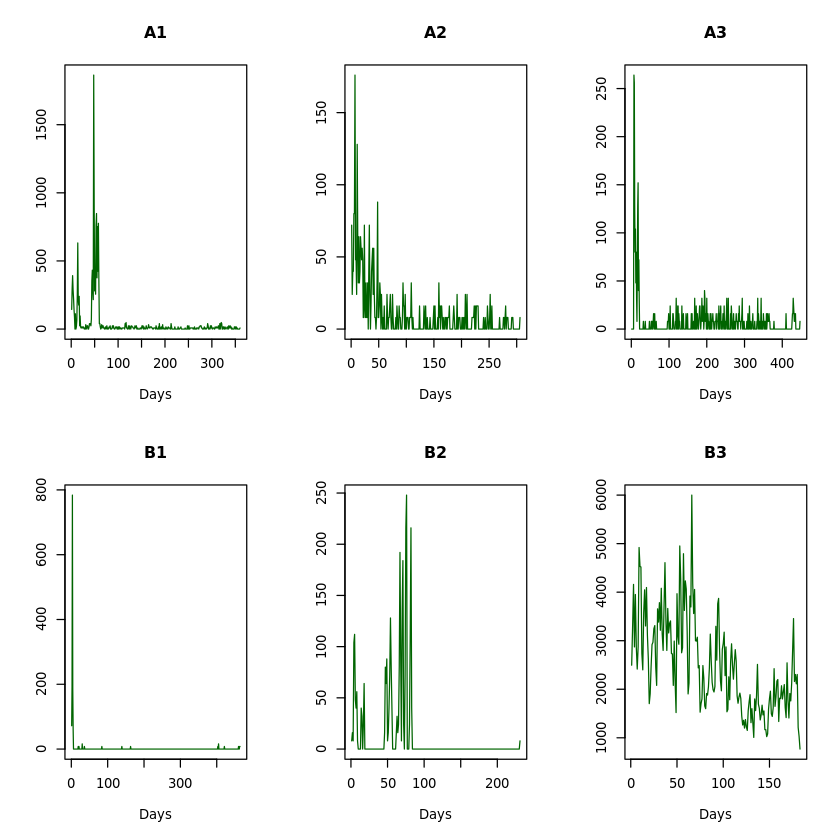}
    \caption{Transactions over Time}
    \label{transfersovertime}
\end{figure}

\begin{figure}[htp]
\centering{}
    \includegraphics[width=8cm,height=6cm]{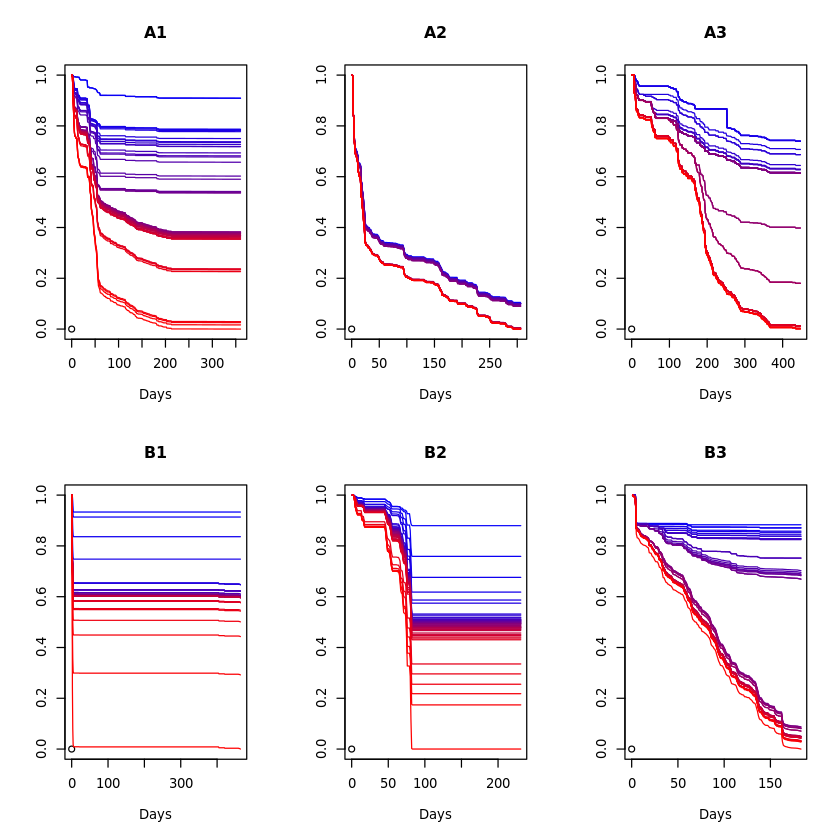}
    \caption{Amount in Play over Time}
    \label{amountinplayovertime}
\end{figure}

\begin{figure}[htp]
\centering{}
    \includegraphics[width=8cm,height=6cm]{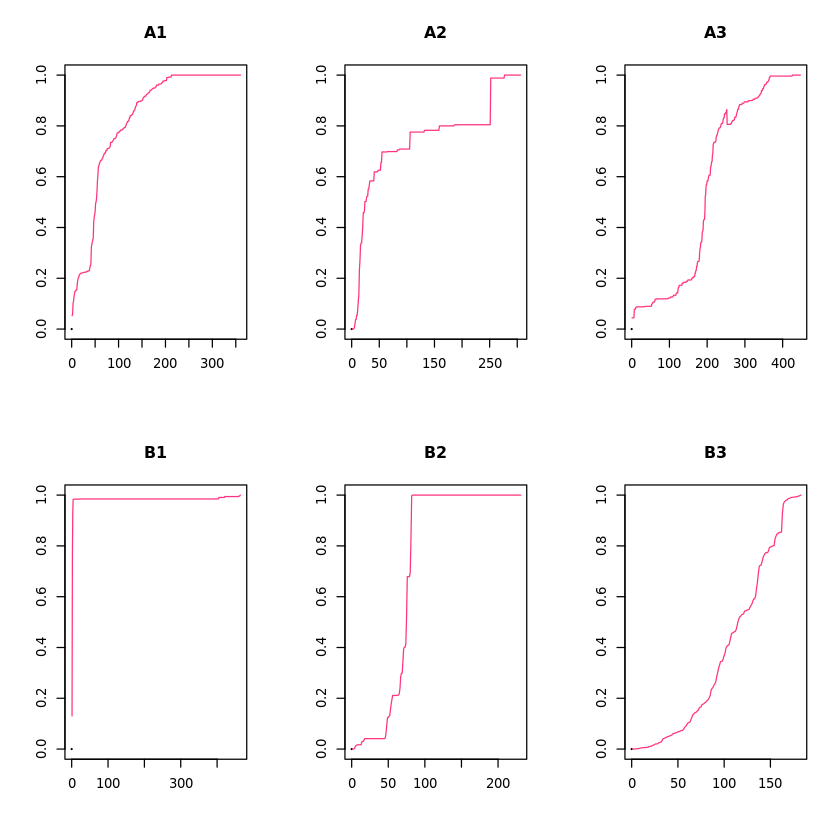}
    \caption{Standard Deviation of Amount in Play over Time}
    \label{stdovertime}
\end{figure}

\begin{figure}[htp]
\centering{}
     \includegraphics[width=8cm,height=6cm]
     {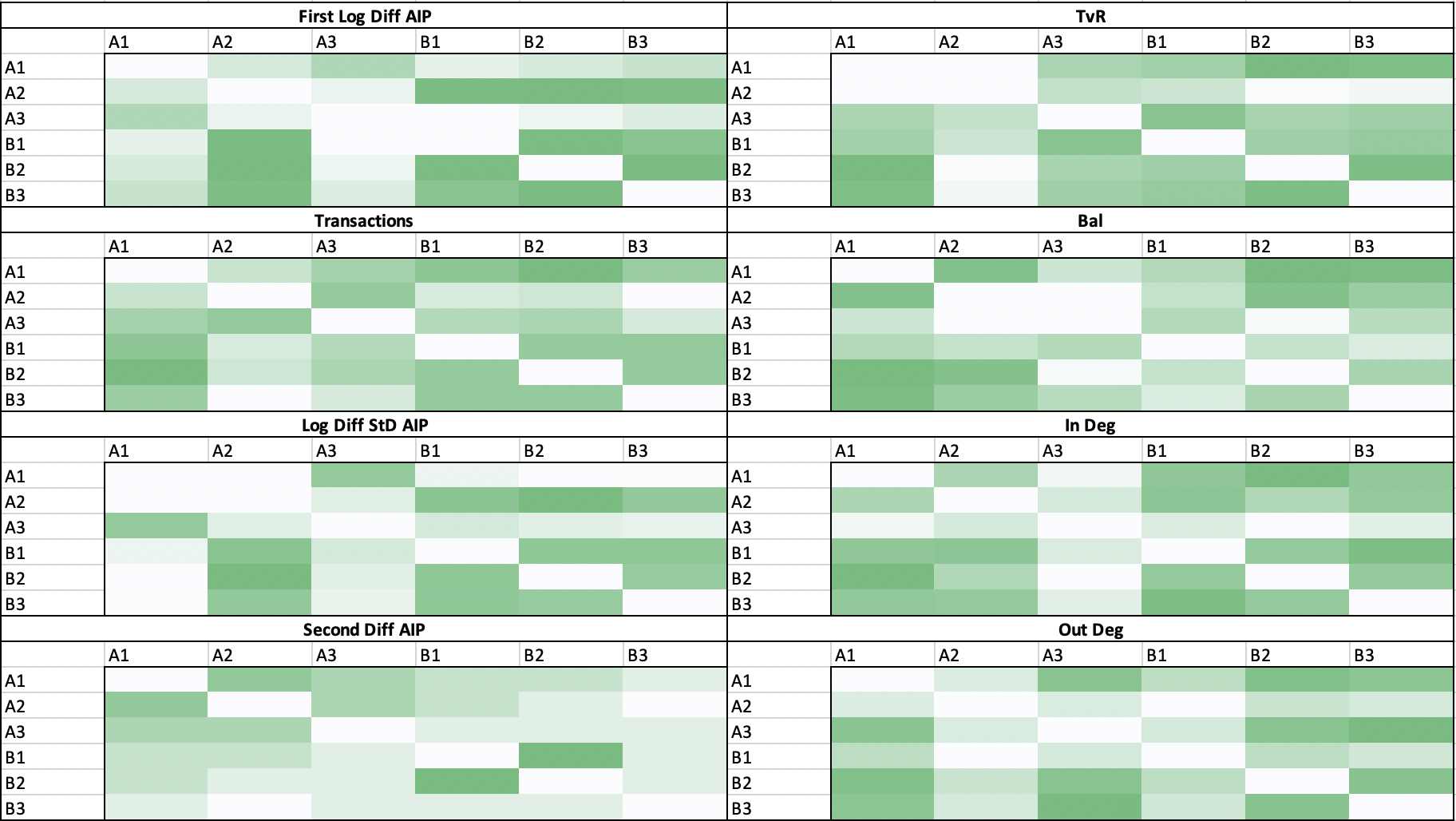}
    \caption{Similarity Matrices of Feature Distributions for Hacking Groups A and B}
    \label{matrices}
\end{figure}

\begin{figure}[htp]
\centering{}
   \includegraphics[width=8cm,height=6cm]
   {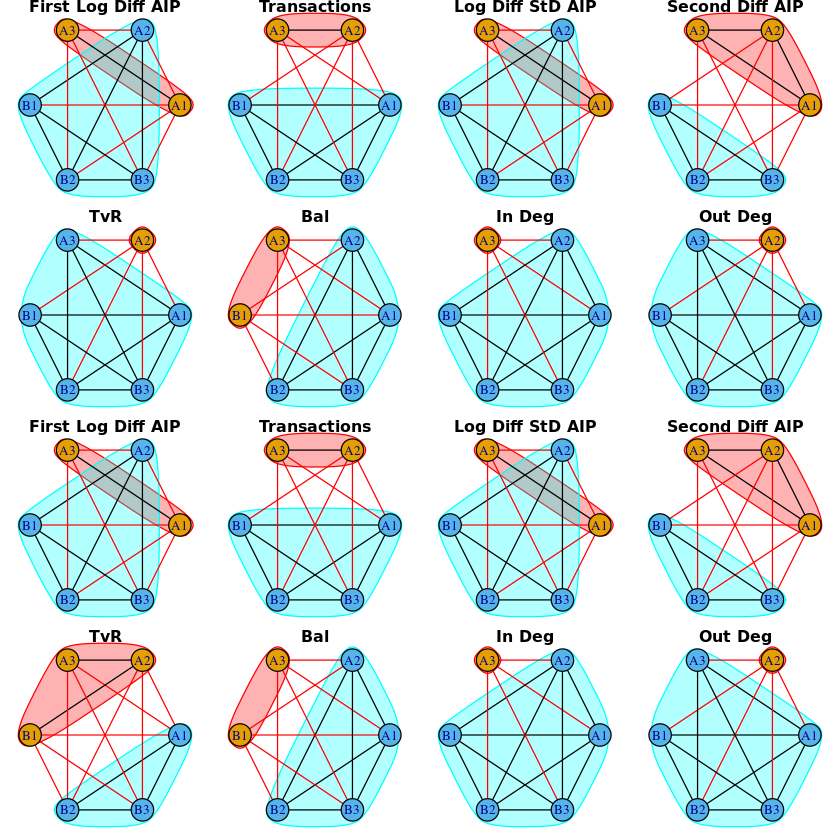}
    \caption{Communities for all features' similarity matrices - First by Walktrap then Modularity Optimization}
    \label{Communities}
\end{figure}
\newpage
As discussed in the Methodology, the communities shown in Figure \ref{Communities} correspond to those identified by two clustering algorithms with the first two rows being Walktrap's output communities on each distribution's similarity network, and the second two rows being the results obtained via Modality Optimization. As can be seen,  similarity matrices derived from different distribution comparisons, whether analyzed by the same or different algorithm lead to different observed communities. Though they are often different, the communities do share some common characteristics with each other. For example, for all but the clustering of Balance similarity and TvR, nodes $\{B1,B2,B3\}$ are always clustered together. Furthermore, $9$ out of the $16$ clusters have at least two members of group A together. 

To better quantify consensus among the results in \ref{Communities}, we first find one node $N$ which remains in the same group through all of the methods (we chose node B6) so as to establish a common group naming (in other words, it is no longer the case that a node is either in the blue or the red group seen in Figure \ref{Communities}, rather that each node is either in the same group as our fixed node or in the opposite group), and then we generate a number $n_{i,j}$ associated to each node $i$ and community $j$, with $j \in \{1,2,3, ... , 16\}$, setting $n_{i,j} = 1$ if $i$ is in the same group as $N$ and $n_{i,j} = 0$ otherwise. We then compute the probability of node $i$ being in the same group as $N$ with $p = \frac{\sum_{j = 1}^{16}n_{i,j}}{16}$. Finally we bisect the vector of values to along its median and obtain the grouping $\{A1,A2,A3\}$, $\{B1,B2,B3\}$. 

This process was repeated using two feature set combinations. The first set contained all 8 features, and its resulting vector was (0.625, 0.5, 0.1875, 0.8125, 1, 1). The second set included only temporal features, namely: LDA, Second Diff(AIP), LDST, and ATVR and had a resulting vector of (0.25, 0.5, 0, 1, 1, 1). Note, that the ground-truth vector is simply (0, 0, 0, 1, 1, 1).
In both cases, the bisection works to successfully find the two communities. In the case of only temporal features, the results are even more compelling where 0.5 can be used to bisect the set of hacks into their respective communities.

\section*{Discussion}

We ran this analysis on historical hacks curated by Chainalysis investigators. The $6$ hacks analyzed were carried out by $2$ distinct and well-known hacking groups. Due to ongoing investigations, the names of the hacking groups cannot be revealed at this time.

Analyzing the subnetworks using our proposed methodology allowed investigators to observe the cash out methods for the different hacking groups. Analyzing each subnetwork based on features above allowed for greater understanding of each specific hack and hacking group, as well as the ability to successfully classify hackers via our pipeline.

\subsection*{Hacking Group Alpha}

 We analyzed three distinct hacks carried out by hacking group alpha. Hacking group alpha is a large, well-funded organization. The hacks analyzed in this paper reveal that the subnetworks tracing funds stolen by hacking group alpha are highly complex, with the stolen funds moving through many nodes. The stolen bitcoins are slowly cashed out through terminal nodes overtime. Investigators confirmed this trend. 

Funds flowing to terminal nodes from the three hacks visualized in Figure \ref{amountinplayovertime} further confirm this trend. Stolen bitcoin being moved by hacking group alpha appear to slowly leak out of possession of the hackers through terminal nodes. Taking both the first and second differences for the amount in play visualized in Figure \ref{amountinplayovertime} demonstrates that the acceleration at which stolen funds exit through terminal nodes is a significant means of clustering the graphs. Just taking first differences successfully clusters hack A3 and hack A1 together. Visually, A1 and A3 are more similar. Looking at the second differences, i.e. the acceleration, for the amount in play visualized in Figure \ref{amountinplayovertime} is most successful at finding communities of hacks. Running community detection on the similarity matrices for the second differences of the amount in play successfully identifies that A1, A2, and A3 belong in the same community. 

The number of transfers that the hackers use to move the funds has also proven significant for helping to effectively classify the hacks according to their hacking groups. As shown in Figure \ref{transfersovertime}, hack A2 and A3 appear to have similar trends in terms of the number of transfers made each day following the hack. The community detection that we ran on the hacks classified these two hacks together when looking only at trends in the frequency of transactions sent to terminal nodes. 

Analysing the variance in the $\rho$ parameter, as visualized by Figure \ref{stdovertime} captures how the share of funds exiting through terminal nodes changes as  $\rho$ approaches 1.  The standard deviation for the  $\rho$ parameter as  $\rho$ approaches 1 approximates the variety in behavior for terminal nodes. Using the log difference in standard deviation across the amount in play by varying $\rho$ allows us to classify hacks A3 and A1 together. Both these hacks had similar changes in the amount in play for each  $\rho$ over time, whereas A2 had some uncharacteristic behavior for hacking group alpha around day 250. A2 was a much smaller sized subnetwork, with only 55 nodes, than A1 and A3, with 1257 and 218 respectively. This made the standard deviation of the amount exiting through terminal node more sensitive as $\rho$ increased. 

We investigated whether the distribution of balances across all the nodes in the hack would be a useful indicator to help classify hacks. This was one of the weakest features used to classify the hacks into hacking groups. As shown in \ref{Distributionlogbalance} there is a wide variety in the distribution across all the nodes in the graph based on their hack balances. Hack A3's distribution, for example, had a higher peak, meaning many of the hacks in A3 held a similar balance. Yet A2 had much more variety across the nodes within the graph in terms of how much stolen bitcoins each node ended up holding. Using the distribution of the log balance by nodes was not useful on its own to help classify hacks, and caused one of the few instances of mistakenly grouping hacks A3 and B4 together as seen in Figure \ref{Communities}.

\subsection*{Hacking Group Beta}

We then analyzed three hacks carried out by the second hacking organization referred to here as hacking group beta. When visualizing the hack subnetworks for hacking group beta, there are striking differences in the cash out mechanisms. Hacking group beta tends to send a majority of its funds through terminal nodes over a short period of time. They tend to sit on their funds quietly, sometimes moving some funds through wallets of interest, but have a characteristically abrupt cash out pattern.

This pattern is visualized in Figure \ref{amountinplayovertime}, where hacks B1, B2 and B3 all have notable vertical drops, representing abrupt moments of cashing out through terminal nodes. Running our community detection algorithms on the first differences of this activity successfully classified all B hacks as belonging together, see  Figure \ref{Communities}, yet also identified hack A2 as fitting a similar pattern. The second differences for the amount in play chart is the best at predicting the proper community assignment. Its top performance can be attributed to its correctly capturing the acceleration of the funds exiting through terminal nodes, which confirms the hypothesis put forward by investigators about temporal trends in exiting funds.

All of the hacks from hacking group beta have a large variance for $\rho$ as $\rho$ approaches one, which can also be visualized in \ref{amountinplayovertime}. This signifies a large range in sending versus receiving behavior for the nodes within the hacking group beta hacks. Funds are exiting through a wide variety of nodes, and not simply hitting one exit point which only ever received funds. 

Looking at the distribution of balances held by the nodes within the subnetwork demonstrates the variety of node behaviors present. However, this was again a weak feature when it came to classifying the hacks through community detection. Hack B1 had many nodes that passed through mixing services which were unclustered in the subnetwork. The mixers would siphon off parts of the stolen funds into consolidator wallets in similar patterns. The investigator only tracked the fattest paths, leaving many of the known nodes passing through mixers with a similar balance. Using balance distribution when the graph is not fully built out was shown not to be useful for community detection. 

We next looked at the variation in the AIP over all $\rho$ as visualized in Figure \ref{stdovertime}. The shape of this graph visualizes how $\rho$ affects the share of funds exiting through terminal nodes. Almost all of hack B1's funds exit through a wide variety of terminal nodes on the first day. The standard deviation peaks at this point, followed by a long period of no fund movements. We successfully classified hacks B1, B2, and B3 together using our community detection algorithms, but hack A2 was mistakenly grouped in when using this feature, as shown in Figure \ref{Communities}.

We then analyzed the number of transactions going to terminal nodes in Figure \ref{transfersovertime}. The number of transactions showed no clear visible pattern to help classify the hacks into hacking groups. While the community detection algorithms successfully classified all three hacks from hacking group beta together, it also picked up hack A1.

\subsection*{Key Takeaways}

We began this analysis by talking with Chainalysis investigators about what they knew about the hacking groups. They indicated that the key differentiation between the two groups, is the pattern by which they hold funds and the subsequent rate at which they cash them out. Our analysis confirms this hypothesis.

We conclude that static features of the charts, such as balance distributions, in degrees, and out degrees are not useful features for classifying the hacks into hacking groups. There are many limitations to these static features. To start, they likely require a fully built out, comprehensive graph. Many of the graphs we chose to analyze were incomplete from the start. This means the takeaways from the static features of the charts were also fundamentally incomplete. 

More importantly, our hypothesis of focusing on the temporal features of the subnetworks, rather than the static features was validated. The results indicate that the patterns by which the subnetworks evolve over time serve as useful features for optimal classification based on the method described in this paper. The optimal classifications in Figure \ref{Communities}, specifically the second difference - or acceleration - of AIP, are most characteristic of the subnetworks temporal nature. Varying $\rho$ to alter our level of resolution into terminal nodes also plays a role in the usefulness of our temporal features and the resulting classifications. The correct classifications were obtained when similarity matrices were built from these temporal features and the community detection algorithms was subsequently run to differentiate the hacking groups based on these features exclusively.

\section*{Conclusion}
Hacks represent an important challenge for law enforcement, the Bitcoin community, and financial institutions. There is opportunity for an algorithmically informed approach to analysis of existing hacks as well as real time monitoring of hacks. This research represents an attempt at building a more rigorous framework for such an approach via an analysis of both the static and temporal features of hack subnetworks and suggests that the temporal features represent an important avenue of exploration for a deeper understanding of the hack subnetworks.

\section*{Future Work}

Due to the small sample size, and the sensitive nature of the underlying data, the tools we develop are currently useful for visualization and description rather than conclusive statistical analysis of existing hacks. We aim to continue gathering data and expand our analysis.


\section*{Acknowledgements}
We thank the Chainalysis investigators for their collaboration.

\section*{Author's contributions}
DG, KG, and YS designed research, performed research, and wrote the paper. All authors read and approved the final
manuscript.

\section*{Funding}
This research was funded by Chainalysis.

\section*{Availability of data and materials}
Due to the sensitivity of the underlying data, we cannot currently release our dataset. 


\end{document}